\documentclass[floats,prd,twocolumn,showpacs,amssymb,nofootinbib]{revtex4}
\usepackage{amssymb}
\usepackage{graphics}
\usepackage{amsmath}
\usepackage{amsfonts}
\usepackage{bm}

\begin{document}

\title{ Cosmic relic abundance and $f(R)$ gravity}%

\author{S. Capozziello$^{a,b}$, M. De Laurentis$^{a,b}$, and G. Lambiase$^{c,b}$}
\affiliation{$^a$Dipartimento di Scienze Fisiche, Universit\'a di Napoli "Federico II", Via Cinthia, I-80126 - Napoli, Italy.}
\affiliation{$^b$INFN Sez. di Napoli, Compl. Univ. di Monte S. Angelo, Edificio G, Via Cinthia, I-80126 - Napoli, Italy.}
\affiliation{$^c$Dipartimento di Fisica "E.R. Caianiello" Universit\'a di Salerno, I-84081 Lancusi (Sa), Italy,}
\affiliation{$^d$INFN - Gruppo Collegato di Salerno, Italy.}
\date{\today}
\def\be{\begin{equation}}
\def\ee{\end{equation}}
\def\al{\alpha}
\def\bea{\begin{eqnarray}}
\def\eea{\end{eqnarray}}

\renewcommand{\theequation}{\thesection.\arabic{equation}}
\begin{abstract}
The cosmological consequences of  $f(R)$  gravity are reviewed in the framework of recent data obtained by PAMELA (Payload for 	Antimatter Matter Exploration and Light-nuclei Astrophysics) experiment. This collaboration has reported an excess of positron events that cannot be explained by conventional cosmology and particle physics, and are usually  ascribed to  the dark matter presence (in particular, weak interacting massive particles). The dark matter interpretation of  PAMELA data has motivated the study of alternative cosmological models (with respect to the standard cosmology) owing to the fact that they predict an enhancement of the Hubble expansion rate, giving rise, in such a way, to thermal relics with a larger relic abundance.
Our analysis shows that  $f(R)$ cosmology allows to explain the PAMELA
puzzle for dark matter
relic particles with masses of the order or lesser than $10^2$ GeV in the regime
$\rho^c \lesssim \rho^m  $ where $\rho^c$ is the curvature density and $\rho^m$ the
radiation density. 
For the  model $f(R)=R+\alpha R^n$, it then
follows that $n\simeq 1$ and small corrections with respect to General Relativity
could lead indeed 
to address the experimental results.  However other interesting  cosmological models can be considered during the pre-BBN epoch as soon as the BBN constraints are relaxed. In such a case, the PAMELA data can be fitted for a larger class of $f(R)$-models.
\end{abstract}


\maketitle

\section{Introduction}
\setcounter{equation}{0}

The discovery of the accelerated expansion of the Universe \cite{accUn} has motivated the developments of several alternative theories of
gravity. These models are built up, typically, either in the framework of the conventional General Relativity (GR) or in the
framework of its possible generalizations or modifications. In the last years, among the different approaches proposed to generalize
Einstein's GR, the so called $f(R)$ gravity has received a growing attention. The reason relies on the
fact that such an approach allows to address the problem of the observed accelerating phase of the Universe, without invoking
exotic sources of dark energy but just extending the standard paradigm of GR. The gravitational Lagrangian for these theories is a generic function of the Ricci scalar
curvature $R$ (not necessarily linear as in the Hilbert-Einstein action), and the corresponding action with the inclusion of standard matter reads
\begin{equation}\label{Lagr}
  S=\frac{1}{2\kappa^2}\int d^4x \sqrt{-g}\, f(R)+S_m[g_{\mu\nu},
  \psi_m]\,,
\end{equation}
where $\kappa^2=8\pi G$. It is a difficult task to deal with higher
order terms in the scalar curvature, thus the forms of $f(R)$ that
have been most  studied in literature are analytic functions of the forms $f(R)\sim
R+\alpha R^n$ or $f(R)\sim R^n$ with $n$ positive or negative. Also broken power laws have been extensively considered
\cite{salvbook,salvreport,odintsovreport,carroll2004,altri,vollick,barraco}.

Current cosmological  observations indicate that our Universe is dominated
by dark matter, responsible of galactic and extragalactic dynamics as well as of structure formations, and dark energy, responsible of the accelerated expansion of the Universe.
In other words, these unknown ingredients mean that we need a source to cluster large scale structure and a source to speed up the Hubble fluid. The ratio between the cold dark matter and the dark energy with the critical density satisfies the bounds \cite{spergel}
 \begin{equation}\label{ratio}
    0.092 \leq \Omega_{CDM} h^2 \leq 0.124\,, \quad 0.30 \leq \Omega_{DE} h^2\leq 0.46\,,
 \end{equation}
where $h=100$Km s$^{-1}$ Mpc$^{-1}$. Although many models have been proposed, the nature of the two components is still unknown.
Favorite candidates for non-baryonic cold dark matter are the so called  WIMPs (weakly interacting massive particles). However, such particles have not been detected yet.

The aim of this paper is to explore the implications of the modified cosmology provided by $f(R)$ gravity to the thermal relic abundance. Alternative cosmologies indeed predict modified thermal histories for relic particles. These modifications, in principle, occur during the {\it pre} big bang nucleosynthesis (BBN) epoch, a period of the Universe evolution not directly constrained by cosmological observations. To account for the enhancement of the expansion rates that cosmological models provide with respect to GR, it is usual to write \cite{fornengo}
 \begin{equation}\label{H=AHGR}
    H(T)=A(T) H_{GR}(T)\,,
 \end{equation}
where $H$ and $H_{GR}$ are the expansion rates of the alternative cosmological model and of GR, respectively. The function $A(T)$ is the  enhancement factor. Its temperature dependence is such that $A(T) > 1$ at large temperature and $A(T)\to 1$ before BBN  set up, i.e. Eq. (\ref{H=AHGR}) holds at early times, while at later time $H=H_{GR}$. This last requirement is necessary due to the successful prediction of BBN on the abundance of primordial light elements. Different cosmological scenarios have been proposed in literature to constrain the function $A(T)$
\cite{randall,kinetion,BD,fornengo,horava,kang,arbey,pallis}.

When the expansion rate of the Universe is enhanced (as compared to that one derived in the framework of GR), thermal relics decouple with larger relic abundance. The change in the Hubble rate may have therefore its imprint on the relic abundance of dark matter, such as WIMPs, axions, heavy neutrinos. This kind of studies is motivated by recent astrophysical results which involve cosmic ray electrons and positrons \cite{PAMELA6,ATIC7,Fermi-LAT8,HESS9-10}, antiprotons \cite{PAMELA11}, and $\gamma$-rays \cite{HESS12,Fermi-LAT13-14}. Particular attention is devoted to the rising behavior of the positron fraction observed in PAMELA experiment \cite{PAMELA6}. Besides the astrophysical interpretation of this phenomenon  \cite{blasi}, it is under investigation  the possibility that the raising of $A(T)$ could be due to dark matter annihilation, dominantly occurred into leptons \cite{arkani,robertson}. In this last case, it is required a large value of $\langle \sigma_{ann} v\rangle$. More specifically, PAMELA and ATIC (Advanced Thin Ionization Calorimeter) data require a cross section of the order  or larger than $\langle \sigma_{ann} v\rangle\sim 10^{26}$ cm$^3$ sec$^{-1}$: in this case thermal relics would have the observed dark matter density  \cite{cirelli}. In this paper, we want to investigate if such data and constraints can be framed in the context of $f(R)$ gravity.

The paper is organized as follows. In Sec. II we derive the  $f(R)$ gravity field equations and discuss the energy-momentum tensor related to the higher-order curvature terms.
In particular, we emphasize the possibility to describe the curvature induced terms as an effective perfect fluid. In Sec. III we write down the expressions
for the {\it energy density} and {\it pressure} induced by higher order curvature-terms for a power law function of the scale factor. We then consider  examples of $f(R)$ models in which the ratio of the {\it curvature energy density} and {\it curvature pressure} is constant. In Sec. IV, we discuss the cosmic enhancement in view of the abundance of thermal relics recalling  the main results reported in  \cite{fornengo}.
Sec. V is devoted to the study of the amplification factor given by (\ref{H=AHGR}) in the framework of $f(R)$ models. The  pre-BBN epoch is considered in Sec. VI. In such a case, $f(R)$-models are less "fine-tuned"  and several of them, in principle, could fit the PAMELA data. 
In Sec. VII we apply
the conformal transformations to recast the $f(R)$  gravity in terms of GR plus a scalar field. The latter gives rise to an effective potential  related to the from of $f(R)$ and its derivative.
Conclusions are drawn in Sec. VIII.
In Appendix  we report some useful  formulas adopted in the main text.

\section{Field equations and  energy-momentum tensor in $f(R)$ gravity}
\setcounter{equation}{0}

The variation of the action (\ref{Lagr}) with respect to the metric yields the field equations
\begin{equation}\label{fieldeqs}
  f' R_{\mu\nu}-\frac{f}{2}\, g_{\mu\nu}-\nabla_\mu \nabla_\nu f'
  +g_{\mu\nu}\Box f'=\kappa^2 T^m_{\mu\nu}\,,
\end{equation}
where the prime indicates the derivative with respect to $R$ and $T^m_{\mu\nu}$ is the energy-momentum tensor of matter.
Here $m$ stands for radiation ($rad$) or standard matter ($mat$). The trace reads
\begin{equation}\label{tracef}
  3\Box f'+f' R-2f=\kappa^2 T^m\,.
\end{equation}
For our aim, it turns out convenient to rewrite (\ref{fieldeqs}) in the form
 \begin{equation}\label{eqcampo2}
    G_{\mu\nu}=\kappa^2 (T^m_{\mu\nu}+T^c_{\mu\nu}) \,,
 \end{equation}
where $G_{\mu\nu}$ is the Einstein tensor
 \[
 G_{\mu\nu}=R_{\mu\nu}-\frac{1}{2}\, g_{\mu\nu} R\,,
 \]
and $T_{\mu\nu}^c$ is the {\it curvature energy-momentum tensor} induced by higher order terms in the curvature invariants  \cite{curvquin}. It is defined as
 \begin{equation}\label{tmunuc}
    \kappa^2 T^c_{\mu\nu}=(1-f')R_{\mu\nu}+\frac{1}{2}(f-R)g_{\mu\nu}+\nabla_\mu \nabla_\nu f'-g_{\mu\nu} \Box f'\,.
 \end{equation}
It gives rise to an {\it effective} description of the source term of Einstein field equations.
A comment is in order. In the right-hand side of Eq.(\ref{eqcampo2})  two effective  fluids  appear indeed: a {\it curvature fluid} and a standard {\it matter fluid}. This representation allows to treat fourth order gravity as standard Einstein gravity in presence of two effective sources \cite{santuzzo}.
This means that such fluids can admit features that could be unphysical for standard matter. Consequently all the thermodynamical quantities associated with curvature should be considered {\it effective} and not bounded by the standard constraints related to matter fields. Moreover, this description does not compromise any of the thermodynamical features of standard matter since Bianchi's identities are separately fulfilled for both fluids as we will show below.

By using the properties
$[\nabla_\gamma, \nabla_\beta]V^\alpha = - R^\alpha_{\,\,\, \rho \beta\gamma} V^\rho$, where $V^\alpha$ is a generic vector, and
$[\nabla_\mu, \nabla_\nu]f'=0$, it is straightforward to show that
 \begin{equation}\label{divTc}
    \nabla_\mu (\kappa^2 T^{c\, \mu\nu})=0\,.
 \end{equation}
As a consequence, the divergences of both sides of Eq. (\ref{eqcampo2}) vanishes, provided
 \begin{equation}\label{divTm}
    \nabla_\mu  T^{m\,\mu\nu}=0\,.
 \end{equation}
The latter equation implies that for $f(R)$  gravity one can separately require that
the energy-momentum tensor of matter is preserved, independently as the gravitational background evolves.

The aspect that arises from the above analysis is that one can define the total energy-momentum tensor as
 \[
 T^{\mu\nu}=T^{m\,\mu\nu}+T^{c\,\mu\nu}\,,
 \]
and then $\nabla_\mu  T^{\mu\nu}=0$. Moreover, it is possible to re-interpret the energy-momentum tensor induced by the curvature
as a perfect fluid (we are assuming a homogeneous and isotropic Universe) \cite{santuzzo}
\begin{equation}\label{tc-perfectfluid}
    T_\mu^{c \,\,\nu}=(\rho^c+p^c)u_\mu u^\nu -p^c \delta_\mu^\nu=(\rho^c, -p^c, -p^c, -p^c)\,,
\end{equation}
where $\rho^c$ and $p^c$ are the {\it energy density} and {\it pressure} induced by the curvature terms;
$u^\mu=(1, 0)$ is the four-velocity of the {\it effective} fluid.   For  matter,  treated as a perfect fluid, one has
\begin{equation}\label{tm-perfectfluid}
    T_{\,\mu}^{m \,\,\nu}=(\rho^m+p^m)u_\mu u^\nu -p^m \delta_\mu^\nu=(\rho^m, -p^m, -p^m, -p^m)\,.
\end{equation}
For a spatially flat FRW's metric
\begin{equation}\label{FRWmetric}
 ds^2=dt^2-a^2(t)[dx^2+dy^2+dz^2]\,,
\end{equation}
one can write the explicit expressions for $\rho^c\equiv \kappa^2 T_{\,\,0}^{c\,0}$ and $-p^c\delta_i^j\equiv \kappa^2 T_{\,\,i}^{c\,j}$ as
 \begin{equation}\label{rhoc}
 \rho^c = (1-f')R_0^0+\frac{1}{2}\, (f-R)-3 H {\dot f}'\,,
 \end{equation}
 \begin{equation}\label{pc}
 p^c = -(1-f')R_i^i-\frac{1}{2}\, (f-R)+{\ddot f}'+2 H  {\dot f}'\,,
 \end{equation}
where
 \[
 {\dot f}'=f'' {\dot R}\,, \quad {\ddot f}'=f''' {\dot R}^2+f'' {\ddot R}\,,
 \]
and $H={\dot a}/a$ is the expansion rate of the Universe.
In the above formula, there is not sum over the indices $i$ in $R_i^i$ (the latter represent the diagonal components of the Ricci tensor), while
the dot stands for the derivative with respect to the cosmic time $t$.

From (\ref{divTc}) one immediately gets
 \begin{equation}\label{eqcontinuityc}
    {\dot \rho^c}+3H(\rho^c+p^c)=0\,.
 \end{equation}
In particular, setting
 \begin{equation}\label{eqstatec}
    p^c = \sigma \rho^c\,,
 \end{equation}
the solution of (\ref{eqcontinuityc}) assumes the standard form
\begin{equation}\label{rhosol}
    \rho^c = \rho_0^c e^{-3\int H (1+\sigma) dt}\,.
\end{equation}
where $\rho^c_0$ is a constant.

In the case in which $\sigma$ is independent of cosmic time, as provided by some models of $f(R)$ gravity, the previous equation reads
\begin{equation}\label{rhosol1}
    \rho^c = \rho_0^c a^{-3 (1+\sigma)}\,.
\end{equation}
Moreover, the equation of continuity for matter gives
\begin{equation}\label{rhomsol}
    \rho^m = \rho_0^m a^{-3 (1+w)}\,,
\end{equation}
where $w=p^m/\rho^m$ is the adiabatic index. We have deliberately indicated the "adiabatic" indices $\sigma$ and $w$ with different symbols in order to point out the different contributions to the cosmic dynamics of the two fluids.

In what follows we will analyze the regime $\rho^c \lesssim \rho^m$, i.e. the effective curvature  is sub-leading with respect to matter. 
As we shall see, the alternative cosmology provided by $f(R)$ gravity 
allows to explain  PAMELA's results  for tiny deviations from GR.
 While the approach to matter fluid is standard, the curvature fluid requires a detailed discussion as discussed in the next section.

\section{Power law expansion of the cosmological background}
\setcounter{equation}{0}

As well known, the high nonlinearity of $f(R)$ gravity makes quite difficult to get exact solutions of field equations \cite{salvbook}.
Here we are interested in finding cosmological scenarios where $\sigma$ is constant in order to compare it with the proper matter adiabatic index $w$.

A possibility to get a constant $\sigma$ is provided by a power law evolution of the scale factor

 \begin{equation}\label{powlawsol}
 a(t)=a_0 t^\beta\,.
 \end{equation}
Such a power law behavior holds in the regime in which we are interested, that is in presence of both curvature and matter  fluids.

Eqs. (\ref{rhoc}) and (\ref{pc}) (see Appendix for details) assume the form
 \begin{equation}\label{rhoc-pw}
 \rho^c =-\frac{\beta}{2(2\beta-1)} R+\frac{f}{2}-\frac{\beta-1}{2(2\beta-1)} f'R -3H{\dot f}'\,,
 \end{equation}
 \begin{equation}\label{pc-pw}
 p^c =\frac{3\beta-2}{6(2\beta-1)} R-\frac{f}{2}+\frac{3\beta-1}{6(2\beta-1)} f'R + {\ddot f}'+2 H {\dot f}'\,,
 \end{equation}
Notice that for $f(R)=R$, both $\rho^c$ and $p^c$ vanish, as expected. Therefore corrections to the Einstein-Hilbert action provide a non-trivial
expression for $\sigma$. In Appendix, we present some details aimed to show that (\ref{eqcontinuityc}) is explicitly fulfilled.

It is worthwhile to write down the explicit expression of $\sigma$ as

 \begin{equation}\label{sigma}
    \sigma = \frac{p^c}{\rho^c}=\frac{1}{3}\frac{{\cal N}}{{\cal D}}\,,
 \end{equation}
\begin{eqnarray}
  {\cal N} &=& (3\beta-2)R-3(2\beta-1)f+(3\beta-1)f'R+  \nonumber \\
    & & \nonumber \\
     & & +\displaystyle{\frac{2(2\beta-3)}{\beta}}f'' R^2 -\displaystyle{\frac{4}{\beta}}f''' R^3\,, \nonumber \\
     & & \nonumber \\
{\cal D} &=& -\beta R+(2\beta-1) f+(1-\beta)f'R-2f'' R^2\,. \nonumber
     \end{eqnarray}
In Table I are reported the expressions of $\sigma$ for some $f(R)$ models.

\begin{table}[t]
\caption{In the Table,
expressions of $\sigma$ for different $f(R)$ models are reported. The quantities $c(\beta, n)$ and $d(\beta, n)$ are constants which depend only on $\beta$ and $n$ (see Eqs. (\ref{c-coeff}) and (\ref{d-coeff})).  These values are independent of the adiabatic index $w$ of matter due to the validity of Bianchi identities  for matter and curvature fluids, which allows to write two independent equations of state.}
\begin{ruledtabular}
\begin{tabular}{ccc}
 & $f(R)$ & $\sigma$  \\  \hline
 1. & $R+\alpha R^2$ & $\displaystyle{\frac{4-3\beta}{3\beta}}$  \\ \hline
  2. & $R+\alpha R^n$ & $\displaystyle{\frac{c(\beta, n)}{3 d(\beta, n)}}$ \\ \hline
 3. & $R+\alpha R^2+\gamma R^3$ & $\displaystyle{\frac{1}{\beta}\frac{\alpha(4-3\beta)+\gamma(4-\beta^2)R}{3\alpha +\gamma (\beta+10) R}}$ \\ \hline
 4. & $\gamma R^n$ & $\displaystyle{\frac{1}{3}\frac{3\beta-2+\gamma c(\beta, n) R^{n-1}}{-\beta  + \gamma d(\beta, n) R^{n-1}}}$ \\
 \end{tabular}
\end{ruledtabular}
\end{table}

Notice that in the cases 1. and 2. of  Table I, the quantity $\sigma$ is constant and independent of $\alpha$, while the case
3. reduces to the case 2. as soon as $\gamma=0$. Moreover, the cases 3. and 4. yield a constant $\sigma$ in the regime of small or large scalar curvature.
In general, $\sigma$ can be zero, negative or positive, depending on the values of the constants $\beta$, $\gamma$, and $n$.

The constants $c(\beta, n)$ and $d(\beta, n)$, appearing in Table I, are defined as
 \begin{equation}\label{c-coeff}
  c(\beta, n)=3(1-2\beta)+(3\beta-1)n+
  \end{equation}
   \[
   +\frac{2n(n-1)}{\beta}[2\beta+1-2n]\,,
   \]
  \begin{equation}\label{d-coeff}
  d(\beta, n)=2\beta-1+n(1-\beta)-2n(n-1)\,.
  \end{equation}

In what follows we shall focalize in particular on $f(R)$ model given by the case 2. in Table I.

\section{The abundance of thermal relics}
\setcounter{equation}{0}

According to standard cosmology and particle physics, the calculation of the relic density of particles is based on the assumption that the
period of the Universe dominated by radiation began before the main production of relics and that the entropy of matter is conserved during this epoch and the successive one. Clearly, a different relic density of particles is expected once these assumptions are modified. In this scenario, therefore, any contribution to the energy density modifies the Hubble expansion rate, which reflects in a modification of the relic density values. Along these lines have been performed investigations considering  Brans-Dicke cosmological models or anisotropic expansions \cite{BD,fornengo}.

The  enhancement function $A(T)$ appearing in Eq. (\ref{H=AHGR}) is conveniently parameterized as \cite{fornengo} (see also \cite{gondolo})
 \begin{equation}\label{A(T)T>Tre}
    A(T)=\left\{ \begin{array}{cc} 1+\eta\left(\frac{T}{T_f}\right)^\nu & \quad  T> T_{re} \\
    1 & \quad T\leq T_{re} \\ \end{array} \right.
 \end{equation}
Here $T_{re}$ denotes the temperature at which the Hubble rate reenters the standard rate of GR (to avoid contradictions with big bang nucleosynthesis, it is required $T_{re}\gtrsim 1$ MeV), while $T_f$ is the temperature at which the WIMPs dark matter freezes out in the standard cosmology ($T_f \simeq 17.3$GeV). Notice that the value of $T_f$, in general, varies by varying the dark matter candidate mass $m_\chi$.  The parameters $\eta$ and $\nu$ are free parameters characterizing the specific cosmological model.  Estimations carried out in \cite{fornengo} have been obtained by setting $T_{re}=1$ MeV.
\textbf{If $\eta\gg1$, the previous equation reads
 \begin{equation}\label{A(T)T>Tre-eta}
 A(T)\simeq \eta\left(\frac{T}{T_f}\right)^\nu\,.
 \end{equation}}
The values of the parameter $\eta$, required to explain the PAMELA data, are
 \begin{equation}\label{etavaluesPAMELA}
    1 \lesssim \eta \lesssim 10^3\,.
 \end{equation}
The corresponding values of the WIMPs masses are in the range
 \begin{equation}\label{massvaluesPAMELA}
10^2 \text{GeV}\lesssim m_\chi \lesssim 10^3\text{GeV}\,.
 \end{equation}
However, for dark matter masses of the order $m_\chi\sim 10^2$GeV, the parameter $\eta$ can also assume values close to zero. Hence
 \begin{equation}\label{etavaluesPAMELAq}
 m_\chi\sim 10^2\text{GeV} \quad \to \quad   0 \lesssim \eta \ll 1\,.
 \end{equation}
These considerations can be framed in the context of $f(R)$ gravity where curvature energy density and pressure play a specific role in the characterization of the
function $A(T)$.

\section{The amplification factor $A(T)$ in $f(R)$ gravity}
\setcounter{equation}{0}

We have now all the ingredients to study the role of  modified cosmology provided by $f(R)$ gravity
to explain the PAMELA data via dark matter relic abundance.

The $0-0$ component of field equations (\ref{eqcampo2}) allows to write the relation between the expansion rates of the Universe
in $f(R)$ cosmology and in the standard cosmology as
 \begin{equation}\label{Hmodified}
    H^2=H_{GR}^2\left(1+\frac{\rho^c}{\kappa^2 \rho^m}\right)\,,
 \end{equation}
where $H_{GR}=(\kappa^2 \rho^m/3)^{1/2}$. Comparing with Eq. (\ref{H=AHGR}), one infers the amplification factor $A(T)$
 \begin{equation}\label{A-f(R)}
    A(T)=\sqrt{1+r}\,,
 \end{equation}
where
 \begin{equation}\label{r-def}
    r\equiv \frac{\rho^c}{\kappa^2 \rho^m}\,.
 \end{equation}
The amplification factor $A(T)$ strictly depends on the ratio between curvature and  matter densities in the field Eq. (\ref{Hmodified}) and, as before discussed,
we will deal with the regime where $\rho^c \lesssim \rho^m$, so that $A(T)\sim 1 + \frac{r}{2}$ (in such a case $A(T)$ is larger than 1 allowing to address the PAMELA result).

The densities are, in principle, functions of the temperature $T$. The conservation of  entropy $S=constant$ implies that
the scale factor and the temperature of the Universe are related by $a=T_0/T$ \cite{kolb} ($T_0=3$K is the present temperature of photons
and we set $a_0=1$, where $a_0$ is the present value of the scale factor).
The combination of Eqs. (\ref{rhomsol}) and (\ref{rhosol}) allows to rewrite Eq. (\ref{r-def}) in the form
 \begin{eqnarray}\label{r-a}
    r &=& \frac{\rho^c_0}{\kappa^2 \rho^m_0} a^{3(w-\sigma)} \\
    &=& \frac{\rho^c_0}{\kappa^2 \rho^m_0}\left(\frac{T_f}{T_0}\right)^{3(\sigma-w)}\times
    \left(\frac{T}{T_f}\right)^{3(\sigma-w)}\,. \nonumber
 \end{eqnarray}
Using the fact that
 \[
\rho^m a^4=\rho_0=\frac{\pi^2 g_*}{30} T_0^4\,,
 \]
where $g_*$ are the relativistic degrees of freedom,  and that
 \[
 h_0^2 \Omega_{0\,rad}=6.15 \times 10^{-5}\,, \quad \Omega_{0\, rad}=\Omega_{0\, \gamma}+\Omega_{0\, \nu}\,,
 \]
one finds
 \[
 \rho_0^m=\Omega_{0\, rad}\rho_{cr}=5.16 \times 10^{-51}\text{GeV}^4\,,
 \]
where $\rho_{cr}=8.4 h_0^2 \times 10^{-47}$GeV$^4$ is the critical density.
Since $\rho^c_0$ must be compared with $\kappa^2 \rho^m_0$, it is worthwhile to report the order of the magnitude of $\kappa^2 \rho^m_0$:
 \[
 \kappa^2 \rho_0^m = 1.3 \times 10^{-87}\text{GeV}^2\,.
 \]

In the regime $r < 1$ ($A\simeq 1+r/2$), Eq. (\ref{A(T)T>Tre}) yields the following expression for the parameter $\nu$ and $\eta$ 
   \begin{equation}\label{eta-nu-f}
    \nu=3(\sigma-w)\,,
    \quad
    \eta=\frac{\rho^c_0}{2\kappa^2 \rho^m_0}\left(\frac{T_f}{T_0}\right)^{3(\sigma-w)}\,.
 \end{equation}
 

In the standard cosmological model, the Einstein field equations admit, for a  radiation dominated Universe, the solution $H_{GR}=1/2t$, i.e. $a= a_0 t^{1/2}$. 
On the other hand, if the time evolution of the cosmological background is governed by field equations (\ref{eqcampo2}), then the scale factor $a\sim t^\beta$ yields
\begin{equation}\label{beta-r}
    H=\frac{\beta}{t}=2\beta H_{GR} \quad \rightarrow \quad \beta=\frac{1}{2}\sqrt{1+r}\,.
\end{equation}
Since $\beta$ is  constant (and positive), the r.h.s. of (\ref{Hmodified}) (and Eq. (\ref{r-a})) must be necessarily independent of the cosmic time,
and therefore on the temperature. Such a condition is fulfilled by requiring\footnote{A comment is in order on Eq. (\ref{sigma=w}). Since our results
are not exacts, we should more properly require $\nu\approx 0$, i.e. $\sigma-w \approx 0$, with  $w = \frac{1}{3}$, $\sigma=w-\varrho$, and $\varrho \ll 1$
This means a very tiny variation of $r$ with $T$. Another possibility to get $r$ constant is to set $\sigma=w=\frac{1}{3}-\varsigma$.
Interactions among massless particles, in fact, lead to running coupling constants, and, hence, the trace anomaly $T^m\propto \beta(g) F^{\mu\nu}F_{\mu\nu}\neq 0$. In Ref. \cite{kitano} it has been studied the thermodynamical potential of a plasma for $SU(N_C)$ gauge theory,
with coupling $g$ and $N_f$ flavors. This study shows that the adiabatic index (for the radiation) is given by $w=\frac{1}{3}-\varsigma$,
where
 \[
 \varsigma=\frac{5}{18\pi^2}\frac{g^4}{(4\pi)^2}\frac{\left(N_C+\frac{5}{4}N_f\right)\left(\frac{11}{3}N_C-\frac{2}{3}N_f\right)}{2+\displaystyle{\frac{7}{2}\frac{N_C N_f}{N_C^2-1}}}\,,
 \]
up to $O(g^5)$ corrections (it is worth noting that typical gauge groups and matter content at very high energies can yield $1-3w \sim 10^{-2} - 10^{-1}$ \cite{kitano}).}
 \begin{equation}\label{sigma=w}
    \nu = 0 \quad \leftrightarrow \quad \sigma-w = 0 \,.
 \end{equation}
In literature, the case $\nu=0$ refers to an overall boost of the Hubble expansion rate.

For $r < 1$, the constant $\beta$ turns out to be related to the parameter $\eta$ by the relation
 \begin{equation}\label{beta-rhoc1}
    \beta=\frac{1+\eta}{2} < \frac{3}{4}\,, \quad\text{as}\quad \eta = \frac{\rho^c_0}{2\kappa^2 \rho^m_0} < \frac{1}{2}\,,
 \end{equation}
However, constraints provided by the big bang nucleosynthesis give a more stringent bound. In fact, in the framework of  $f(R)$ cosmology one gets \cite{pizza}
 \begin{equation}\label{betaBBN}
    \beta \lesssim \frac{1}{2}+ 10^{-3}\,,
 \end{equation}
so hat
 \begin{equation}\label{eta-BBN}
    \eta\lesssim 2 \times 10^{-3}\,.
 \end{equation}
See also \cite{kolb,bernstein} for a general discussion on big bang nucleosynthesis.

The previous results have a general validity because they do not refer to a specific form of $f(R)$. As a consequence, $f(R)$ cosmology represents a suitable framework 
for getting the amplification of the expansion rate of the Universe, and therefore to explain the PAMELA data.
From Eq. (\ref{etavaluesPAMELAq}), we therefore conclude that the mass of the relic dark particles is $\sim 10^2$GeV.

In what follows, we shall apply the above results to the model given by
 \begin{equation}\label{fmodel}
f(R)=\omega R + \alpha R^n\,,
 \end{equation}
where we insert the constant $\omega$ to figure out terms coming from GR (at the end it will be set equal to 1).

Let us start by checking the consistency  between Eqs. (\ref{rhoc}) and (\ref{rhosol}) for this model.
From Eq. (\ref{rhoc}), it follows
\begin{eqnarray}\label{rhoc-n}
    \rho^c &=& (\omega-1)\frac{\beta R}{2(2\beta-1)}+ \\
         & & +\frac{\alpha R^n}{2(2\beta-1)}[(2-n)\beta -(2n-1)(n-1)] \nonumber\\
&=& \frac{\rho^c_0}{a^{2n/\beta}} \nonumber\,.
\end{eqnarray}
Here $\rho^c_0$ is defined as
\begin{equation}\label{rhoconstat}
    \rho^c_0 \equiv \alpha \Gamma\,,
\end{equation}
with
 \begin{equation}\label{Gamma}
    \Gamma\equiv \frac{(2-n)\beta -(2n-1)(n-1)}{2(2\beta-1)}\, [-6\beta (2\beta-1)]^n a_0^{2n/\beta}\,.
 \end{equation}
The last expression has been obtained for $\omega =1$.
In the above equations we used $R=-6\beta(2\beta-1)/t^2$ and $t=(a/a_0)^{1/\beta}$.

By comparing (\ref{rhoconstat}) with (\ref{rhosol1}), one gets
 \begin{equation}\label{equality}
    \beta=\frac{2n}{3(\sigma+1)}\,.
 \end{equation}
It is easy to verify that using $\displaystyle{\sigma=\frac{c(\beta, n)}{3d(\beta, n)}}$ (case 2. in Table I), with $c(\beta,n)$ and $d(\beta, n)$ defined in (\ref{c-coeff}) and (\ref{d-coeff}), respectively, Eq. (\ref{equality}) is automatically fulfilled.

In the case in which $\sigma\approx w \approx 1/3$, Eq. (\ref{equality}) gives
\begin{equation}\label{equalitynmezzi}
 \beta \approx \frac{n}{2}\,.
\end{equation}

Equations (\ref{beta-rhoc1}), (\ref{eta-BBN})  and (\ref{equalitynmezzi}) imply
\begin{equation}\label{n-R2-BBN}
n = 1+\eta \lesssim 1+ 2\times 10^{-3}\,, \quad \beta \simeq \frac{1}{2}\,.
\end{equation}
Hence, the model  works provided the constraint (\ref{n-R2-BBN}) is satisfied.
Such a result agrees with Solar System constraints \cite{tsuji} and the range in Eq.(\ref{etavaluesPAMELAq}).

\section{Analysis  in the pre-BBN epoch}  

Let us now study the case in which the BBN-constraint (\ref{betaBBN}) is relaxed.  This means that we are well before the BBN-epoch. This case is justified by the fact that, in the model proposed in  \cite{fornengo}, the enhancement function $A(T)$ is set to 1 as soon as the Universe reaches the BBN era. From this instant, it is assumed to evolve according to GR. In this perspective, it turns out interesting to analyze  the case  $ r \gtrsim 1$ where cosmological models are less fine-tuned.

We can take into account  the  case
 $r \gtrsim 1$  where $A \simeq \sqrt{r}/2$.  From Eq. (\ref{A(T)T>Tre-eta}) one gets that the parameters $\nu$ and $\eta$
  assume the form
     \begin{equation}\label{eta-nu-f2}
    \nu=\frac{3(\sigma-w)}{2}\,,
    \quad
    \eta=\left(\frac{\rho^c_0}{2\kappa^2 \rho^m_0}\right)^{1/2}\left(\frac{T_f}{T_0}\right)^{3(\sigma-w)/2} \,.
 \end{equation}
  The constant $\beta$ turns out to be related to the parameter $\eta$ by the relation
 \begin{equation}\label{beta-rhoc}
    \beta=\frac{\sqrt{\eta}}{2}\gtrsim \frac{1}{2} \,, \quad \text{as}\quad \sqrt{\eta} = \left(\frac{\rho^c_0}{\kappa^2 \rho^m_0}\right)^{1/2}\gtrsim 1\,,
 \end{equation}
which means that the relic dark particles have masses $> 10^2$GeV, see Eq. (\ref{etavaluesPAMELA}).

 Combining (\ref{beta-rhoc}) with (\ref{equalitynmezzi}) it follows that in the regime $r \gtrsim 1$, the parameter $n$ has to satisfy the condition
 \begin{equation}\label{condition-n}
 n = \sqrt{\eta} \, \gtrsim 1\,.
 \end{equation}
Therefore, the model (\ref{fmodel})  solves the PAMELA puzzle as soon as the exponent
$n$  assumes values greater than 1, while the Universe evolves as $a\sim t^\beta$, with $\beta \gtrsim 1/2$.


Clearly, looking at the Table I, in the regime $R\gg 0$, the parameter $\sigma$ is a constant whose signature depends on the values of $\beta$ and $n$.
For example, considering the Case 3. in  Table I gives 
\begin{equation}
\sigma=\frac{1}{\beta}\left(\frac{4-\beta^2}{\beta+10}\right)\,.
\end{equation} 
This means that, for large $\beta$, $\sigma$ approaches to -1 and then the curvature fluid behaves as a cosmological constant (see Eq.(\ref{rhosol1})).
In conclusion, it is possible to show that relaxing the BBN-constrain, i.e. during the pre-BBN era, there are several modified gravity models (in particular $f(R)$-models) capable of fitting the PAMELA data.

\section{Effective potentials in the Einstein frame}

The further gravitational degrees of freedom coming from $f(R)$ gravity can be figure out as additional scalar fields.  In this case, the  enhancement necessary to explain the relic abundance (i.e. the amplification factor $A(T)$) can be tracked by the scalar field dynamics.

Let us  discuss the conformal transformation of   $f(R)$ gravity which can be transformed into
 Einstein's gravity  plus a minimally coupled scalar field \cite{salvbook,odintsovreport}. We can assume that the energy density $\rho^c$ and
the pressure $p^c$ are dominant with respect to matter. This assumption is very natural since curvature components can play the role of cosmological dark energy which is larger than matter component \cite{curvquin}. Furthermore, the assumption is justified by the fact that we are in the regime $r\gtrsim 1$.  To better appreciate the role of the scalar field when passing to the Einstein frame, we can neglect its coupling to matter fields since it is not particularly relevant in this epoch and its signature can be hardly observed. 

Let us consider the conformal transformation on the metric
$g_{\mu\nu}$
 \begin{equation}\label{conftransf}
    {\tilde g}_{\mu\nu} = e^{2\chi}g_{\mu\nu}\,.
 \end{equation}
By choosing
 \begin{equation}\label{chidef}
    \chi=\frac{1}{2}\ln |f'(R)|\,,
 \end{equation}
and setting
\begin{equation}\label{varphi}
    k \varphi = \chi\,, \qquad k=\frac{1}{\sqrt{6}}\,.
\end{equation}
it can be shown that the Lagrangian density of $f(R)$ in (\ref{Lagr}) can be recast in the (conformally) equivalent form \cite{salvbook,salvreport,odintsovreport}
\begin{equation}\label{euivform}
    \sqrt{-g}f(R)=\sqrt{-{\tilde g}}\left(-\frac{1}{2}{\tilde R}+\frac{1}{2}\nabla_\mu \varphi \nabla^\mu \varphi - V\right)\,,
\end{equation}
while the field equations reads
 \begin{equation}\label{einsteintransf}
    {\tilde G}_{\mu\nu}=\kappa^2 \left[\nabla_\mu \nabla_\nu-\frac{1}{2}{\tilde g}_\mu\nu \nabla_\rho \varphi \nabla^\rho  \varphi+{\tilde g}_{\mu\nu}
    V(\varphi)\right]\,.
 \end{equation}
The potential $V$ is defined as
\begin{equation}\label{potV}
    V= \frac{f-R f'}{2f^{'\, 2}}\,.
\end{equation}
Let us figure out the explicit form of the potential $V$ in the case in which $f=R+\alpha R^n$. Inverting (\ref{chidef}), one gets
\begin{equation}\label{chiinvet}
    f' = e^{2k \varphi}\,,
\end{equation}
Using $f'=1+\alpha n R^{n-1}$, the previous equation yields
 \begin{equation}\label{Rvsphi}
    R=\left[\frac{1}{\alpha n}(e^{2k\varphi}-1)\right]^{\frac{1}{n-1}}\,.
 \end{equation}
The potential (\ref{potV}) reads
\begin{eqnarray}
  V &=& \frac{\alpha(1-n)}{2}e^{-4k\varphi}\left[\frac{1}{\alpha n}(e^{2k\varphi}-1)\right]^\frac{n}{n-1} \label{V1} \\
   &=& \frac{2^{\frac{1}{n-1}}\alpha (1-n)}{(\alpha n)^{\frac{n}{n-1}}} e^{k \frac{4-3n}{n-1} \varphi}\left[\sinh k\varphi\right]^{\frac{n}{n-1}}\,.
   \nonumber
\end{eqnarray}
For $k\varphi\ll 1$ the potential assumes a power law behavior
\begin{equation}\label{potV2}
    V\simeq V_0 \varphi ^\delta\,,
\end{equation}
where
 \[
 V_0 \equiv \frac{2^{\frac{1}{n-1}}\,\alpha (1-n)}{(\alpha n)^{\frac{n}{n-1}}}\,,
 \quad \delta \equiv \frac{n}{n-1}\,.
 \]
Such a form of potential has been widely studied in literature in the framework of alternative theories
of gravity \cite{salvreport}.

\begin{figure}
\resizebox{7.5cm}{!}{\includegraphics{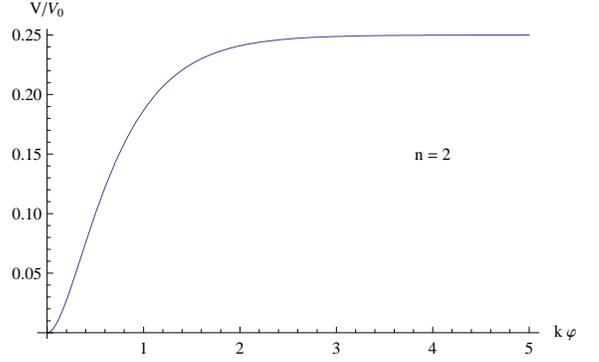}}
\caption{$V/V_0$ vs $k\varphi$ for $n=2$. $V/V_0\to 1/4$ as $k \varphi \gg 1$.}
 \label{fig1}
\end{figure}

\begin{figure}
\resizebox{7.5cm}{!}{\includegraphics{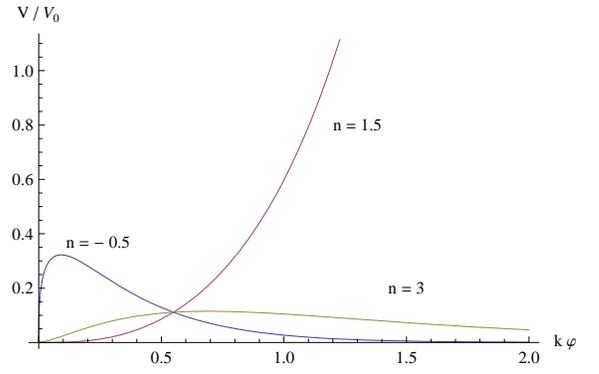}}
\caption{$V/V_0$ vs $k\varphi$ for $n=-0.5$, $n=1.5$, and $n=3$. for large $k \varphi$, the potential approaches to zero for $n=-0.5$ and $n=3$,
while grows up for $n=1.5$.}
 \label{figall}
\end{figure}

It is worthwhile to point out that, in the case $n=2$, the potential (\ref{V1}) tends to a constant value for large $k\varphi$
 \begin{equation}\label{Vconstant}
    V \to \frac{V_0}{4} \qquad\quad \text{as}\,\,\,\, k\varphi \gg 1\,.
 \end{equation}
Therefore, in this regime the potential plays the role of a cosmological constant.
This behavior is shown in Fig. \ref{fig1}, where it is plotted $V/V_0$ vs $k\varphi$ for $n=2$.
To compare with other values of $n$, we also plot $V/V_0$ for $n=-0.5$, $n=3$ and $n=1.5$, see Fig. \ref{figall}.
In the first two cases, the potential $V$ approaches to zero as $k \varphi \gg 1$,
in the latter $V$ grows for increasing $k \varphi$. 

Specifically,  the amplification factor (\ref{A-f(R)}) can be expressed in terms of the scalar field $\varphi$. 
By using the scale factor (\ref{powlawsol}) and the relation between the Ricci curvature and the scalar field, Eq. (\ref{Rvsphi}),
one obtains
 \begin{equation}\label{a-phi}
    a(\varphi)=a_0 |6\beta(2\beta-1)|^{\frac{\beta}{2}}(\alpha n)^{\frac{\beta}{2(n-1)}}(e^{2k\varphi}-1)^{-\frac{\beta}{2(n-1)}}\,.
 \end{equation}
As a consequence, the amplification factor $A(\varphi)$ reads
 \begin{equation}
 A(\varphi)=[1+r(\varphi)]^{1/2}\,,
 \end{equation}
where the quantity $r$, defined in (\ref{r-a}), is given by
 \begin{equation}\label{r-phi}
 r(\varphi)= r_0\, \left[2^{2k\varphi}-1\right]^{-\frac{3\beta(w-\sigma)}{2(n-1)}}\,,
     \end{equation}
with
 \begin{equation}
 r_0 \equiv \frac{\rho_0^c}{\kappa^2 \rho_0^m} \left[a_0 |6\beta(2\beta-1)^{\frac{\beta}{2}} (\alpha n)^{\frac{\beta}{2(n-1)}}\right]^{3(w-\sigma)}\,.
 \end{equation}
 These considerations mean that it is straightforward to translate results in Sec.V in the Einstein frame and the amplification factor, being linked to the Hubble parameter, strictly depends on the form of the potential $V(\varphi)$ and then $\varphi$. In other words, dynamics related to $\rho^c$ and $p^c$ can be recast in terms of $V(\varphi)$ and $\varphi$.
 
  Reversing the argument,  the PAMELA data (an then the requested amplification factor to explain them) could be a probe for alternative theories of gravity, in particular for $f(R)$ gravity.
  
Finally, as discussed in \cite{troisi, moruno,moruno1}, it is worth noticing  that $f(R)$ gravity can, in general, give rise to cosmological viable models compatible with radiation and  matter-dominated epochs evolving into  late accelerated phases.  This means that reliable models  can be phenomenologically reconstructed by means of observational data.

\section{Conclusions}

There is nowadays a great interest toward the recent data of PAMELA experiment since they could represent a possible signature for dark matter. Among the various mechanisms that have been proposed to explain the PAMELA results, the models based on alternative cosmologies represent suitable candidates. These models rely on the fact that allow an enhancement of the early expansion rate of the Universe, hence large annihilation cross sections. At the same time, such models are  also compatible with other observations \cite{salvbook}.

In this paper, we faced the problem of the PAMELA puzzle in the context of cosmological models provided by $f(R)$ gravity.
Our approach is based on the effective description of the energy-momentum tensor (the total energy-momentum tensor is written as the sum of the energy-momentum tensor of ordinary matter, $T_{\mu\nu}^m$, plus the effective energy-momentum tensor $T_{\mu\nu}^c$ induced by curvature terms appearing in the nonlinear action of gravity - these two fluids satisfy separately Bianchi's identities) and on the assumption that the scale factor evolves as a power law ($a\sim t^\beta$). By confining ourselves to the case in which $T_{\mu\nu}^m \gtrsim T_{\mu\nu}^c$, we found that value of the dark matter masses necessary to explain the PAMELA puzzle turn out to be of the order $m_\chi \sim 10^2$GeV. In particular, for the model $f(R)=R+\alpha R^n$ it follows that to explain the PAMELA data, the admissible value of the exponent $n$ is given by Eq. (\ref{n-R2-BBN}) at the BBN epoch. Relaxing such a hypothesis and considering also the pre-BBN epoch, the model is less "fine-tuned" and other $f(R)$ models agrees with data. The evolution can be tracked considering conformal transformations. In the Einstein frame, the amplification factor $A$ depends on a scalar whose potential determines the model dynamics. 

The analysis performed in this paper relies on $f(R)$ models assumed as function of  the scalar curvature $R$. However, other curvature invariants
like Riemann and Ricci invariants can be considered to refine the analysis.  As final comment, it is important to stress that PAMELA data are very strict results indicating that revisions are necessary at fundamental physics  and cosmological levels.

\appendix

\section{ Power law solutions, continuity and field equations}
\setcounter{equation}{0}

In this Appendix we provide details of formulas used in the main text. For the FRW metric and for a scale factor evolving as a power law,
$a=a_0 t^\beta$, we have
 \begin{equation}
 \frac{\dot a}{a}=H=\frac{\beta}{t}\,, \quad \frac{\ddot a}{a}=\frac{\beta(\beta-1)}{f^2}\,,
 \end{equation}
 \begin{equation}
 R_0^0=-3({\dot H}+H^2)=-\frac{3\beta(\beta-1)}{t^2}=\frac{1}{2}\frac{\beta-1}{2\beta-1}\, R\,,
 \end{equation}
 \begin{equation}
 R_i^i=-({\dot H}+3H^2)=-\frac{\beta(3\beta-1)}{t^2}=\frac{1}{6}\frac{3\beta-1}{2\beta-1}\, R\,,
 \end{equation}
 \begin{equation}
 R=-6({\dot H}+2H^2)=-\frac{6\beta(2\beta-1)}{t^2}\,,
 \end{equation}
 \begin{equation}
 {\dot R}=-6({\ddot H}+4H{\dot H})=-\frac{2R}{t}\,,
 \end{equation}
 \begin{equation}
 {\ddot R}=-6({\dddot H}+4{\dot H}^2+4H{\ddot H})=\frac{6R}{t^2}\,,
 \end{equation}
 \begin{equation}
 \Box f'={\ddot f}'+3H{\dot f}'\,,
 \end{equation}
 \begin{equation}
  \nabla_0 \nabla_0 f'= {\ddot f}'\,,  \quad \nabla_i \nabla^j f'=H {\dot f}' \delta_i^j\,.
 \end{equation}

Let us now  show that Eq. (\ref{eqcontinuityc}) is fulfilled assuming that the Universe evolves
with a scale factor of the form $a\sim t^\beta$. By using the relation ${\dot R}=-2R/t=-2HR/\beta$, one can write
${\dot \rho}^c$ as
 \begin{equation}
 {\dot \rho}^c=-\frac{HR}{2\beta-1}(-1+f'-f'' R)-3H{\ddot f}'\,.
 \end{equation}
On the other hand, summing up (\ref{rhoc-pw}) and (\ref{pc-pw}) and using $H{\dot R}=R^2/3(2\beta-1)$ it follows
 \begin{equation}
 \rho^c+p^c = \frac{R}{3(2\beta-1)}\, (-1+f'-f'' R)+{\ddot f}'\,.
 \end{equation}
These equations show that Eq. (\ref{eqcontinuityc}) is fulfilled.

Finally,  we discuss in some details the solutions of the field equations (\ref{fieldeqs}) in the regime
$r \lesssim 1$. For the (\ref{FRWmetric}) they read 
 \begin{equation}
-3\frac{\ddot a}{a}f'-\frac{f}{2}+3\frac{\dot a}{a}f'' {\dot R}=\kappa^2 \rho^m\,,
\end{equation}
\begin{equation}
\left(\frac{\ddot a}{a}+2\frac{{\dot a}^2}{a^2}\right)f'+\frac{f}{2}-2\frac{\dot a}{a}f''{\dot
R}-f'''{\dot R}^2-f'' {\ddot R}=\kappa^2 p^m\,,
\end{equation}
For the model (\ref{fmodel}) and $a=a_0 t^\beta$, the above equations can be recast in the following form
\begin{equation}
  \frac{2\beta}{2\beta-1} \omega R+\frac{\alpha \,d(\beta, n)}{2(2\beta-1)}\, R^n = \kappa^2 \rho^m\,, \label{C1}
  \end{equation}
  \begin{equation}
  \frac{3\beta-2}{6(2\beta-1)} \omega R+\frac{\alpha \, c(\beta, n)}{6(2\beta-1)}\, R^n = \kappa^2 w \rho^m\,, \label{C2}
\end{equation}
where we have used the above results  and $p^m = w \rho^m$ ($w=1/3$), while $c$ and $d$ are defined by Eqs. (\ref{c-coeff}) and (\ref{d-coeff}).

In the regime $r< 1$, therefore the $R^n$-term is a perturbation, Eqs.  (\ref{C1}) and (\ref{C2}) show that a possible solution
is provided by $n \sim 1$, according to (\ref{n-R2-BBN}).

\end{document}